\documentclass[12pt]{article}
\usepackage{amsmath}
\usepackage{amssymb}
\usepackage{graphicx}
\usepackage{epsfig}
\newcommand{\bi}[1]{\bibitem{#1}}

\newcommand{\be}{\begin{eqnarray}}
\newcommand{\ee}{\end{eqnarray}}
\newcommand{\ra}{\rightarrow}

\newcommand{\chapter}[1]{\tmsection{\begin{center}\huge #1\end{center}}}

\begin{document}
\baselineskip=14.5pt
\pagestyle{plain}
\begin{titlepage}

\begin{flushright}
hep-th/0204038\\
HUTP-02/A007\\
IFIC/02-13\\
\end{flushright}

\begin{center}
{\huge Entropy-Area Relations in Field Theory}\\ [8pt]
\end{center}

\vskip 0.4cm
\begin{center}
{\large Lisa Randall$^1$, Ver\'onica Sanz$^2$, Matthew D. Schwartz$^1$}\\

\vskip 1cm
$^1$ Jefferson Physical Laboratory\\
 Harvard University, Cambridge, MA 02138\\
 $^2$ Depto. de F\'\i sica Te\'orica - IFIC, Centro Mixto \\
 Universidad de Valencia-CSIC, Valencia, Spain\\

\end{center}

\vskip 1.2cm
\begin{center}
{\large Abstract}
\end{center}
\noindent 
We consider the contribution to the entropy from
fields in the background of a curved time-independent
metric. To account for the
curvature of space, we postulate a position-dependent
UV cutoff. We argue that a UV cutoff on energy
naturally implies an IR cutoff on distance.
With this procedure, we calculate the scalar contribution in
a background
anti-de Sitter space, the exterior of a black hole, and de Sitter space.
In all cases, we find results that can be simply interpreted
in terms of local energy and proper volume, yielding insight into
the apparent reduced dimensionality of systems with gravity.

\vskip 0.4cm
{\vskip 5pt \footnoterule\noindent
{\footnotesize  {\tt randall@physics.harvard.edu}}\\
{\footnotesize  {\tt Veronica.Sanz@ific.uv.es}}\\
{\footnotesize {\tt matthew@feynman.princeton.edu}}}

\end{titlepage}
\newpage

\section{Introduction}
The Bekenstein bound severely restricts the number of degrees of
freedom in a gravitational system, bounding the entropy by
$S<\frac{\mathcal{A}}{4 \pi} M^2_{\mathrm{Pl}}$,
where ${\mathcal A}$ is the area of the system. 
We would like to understand how to formulate field theory so that
it manifestly reflects this lower number of degrees of freedom.
The holographic principle tells how to associate a bulk region of space
with a boundary region \cite{thholo,sussholo}. 
In the case of AdS/CFT,  the holographic
principle is explicit because there exists an explicit duality between
bulk and boundary theories \cite{maldacena}. 
Detailed tests of the conjecture have
been performed \cite{maldacena,witten,gubser} and new principles
have been conjectured based on this correspondence
\cite{polsussto}. 
Most systems will not 
exhibit this correspondence so simply; at the very least, the holographic
dual to a bulk system would generally not be a local field theory, 
as with the linear
dilaton theory for example \cite{ldt}.  
Nonetheless, because AdS is so well explored,
it makes sense to extrapolate lessons about the field theory in the bulk
to see how far one can get by studying only the bulk theory, and not
using anything about the existence of a holographic dual 
theory.
This is motivated by the calculation done in Ref. \cite{us1,us2}, where we 
derived a logarithmic running of the coupling in a {\it five-dimensional}
calculation. We will use the regularization procedure employed there
based on a cutoff on local energy
to extract the essential physics that led to the appearance of reduced
dimensionality of the field theory.  Of course, for AdS space, the
apparent reduced dimensionality might be simply the fact that volumes
behave like areas, in
the sense that integration over the direction perpendicular to the
boundary only multiplies the area by a finite length, the AdS
curvature, independent of the size of this dimension.
In fact, this is in some sense true, and it is the purpose of this
paper to generalize this simple intuition.

We are led to ask if we calculate the number of
degrees of freedom in a curved space by generalizing
this method, do we always get the right
answer (or at least the correct dimensionality describing the
number of degrees of freedom)  by
generalizing our regularization procedure.
Clearly, because the cutoff is an artifact of not
doing a full string theory calculation, we will not obtain precise
answers  with a trustworthy numerical coefficient.  Nonetheless,
it is of great interest that we will find that
our answers scale in a way consistent
with the Bekenstein formula; that is, the degrees of freedom fit
on a space of one lower dimension. The main purpose of this paper
is to demonstrate that by eliminating the region of space-time
where energies are above a local cutoff scale, we can understand
intuitively why and how a theory becomes holographic, in the sense
of having degrees of freedom reflecting a space of reduced
dimensionality.  We will show this for global AdS, de Sitter
space, and black holes.

Our procedure incorporates a cutoff on local energy that is
reinterpreted as a cutoff on position for a given energy.
Now we make a further assumption, that may be interpreted as a weak
form of a UV/IR correspondence, and assume that there is an
energy-dependent cutoff on the {\it size} of the space (IR)
that reflects the energy cutoff (UV). That
is, we assume that the region which does not permit the given
value of energy does not exist and we quantize the system on this
space of reduced size.   This procedure
will be explained more fully in Section \ref{secOTH} but clearly
relies on an IR cutoff on the {\it size} of the space resulting
from a UV cutoff on the local energy.
The intuition for this assumption comes
from the field theory AdS calculation of Ref. \cite{us2}.
It reflects the fact that the space probed by the high
energy modes is effectively smaller than the full space.
The low mass modes have substantial amplitude only
outside this region. Furthermore, the coupling
changed in such a way to compensate for the larger
mode spacing.
Without detailed knowledge of the Kaluza-Klein spectrum,
one cannot in general deduce the reduced dimensionality through the
energy-dependent cutoff alone. Imposing the IR cutoff that
reflects the local UV cutoff reproduces this structure of the modes.

This picture gives a simple origin for the reduction of degrees of
freedom.  By forcing local energy,
$\sqrt{g^{t t}} E$, to be less than a cutoff $\Lambda $, the high energy
modes will in general probe less of the space than the low energy
modes.
The spatial cutoff implements the correct counting. It
reflects the physical fact that at high energy, there are regions
of space that are inaccessible, so effectively the boundary depends on energy.
This reduction of volume at high energy can have the same
effect on thermodynamic quantities as confining the theory to a
subspace, for example, the boundary. There is not necessarily a
boundary theory; however, in certain cases (when a horizon exists
for example), the physics is dominated by the degrees of freedom
concentrated near this region.

Let us summarize the similarities and differences between our
approach, and one that postulates the existence of a boundary
theory. In both approaches, we find in certain known examples that
the number of degrees of freedom scales with area, not volume. In
our description, the degrees of freedom reside throughout the
space, but might be peaked on the boundary. This should be
contrasted with a boundary description where the degrees of
freedom are fundamental degrees of freedom  on a holographic
screen. Our description only involves bulk degrees of freedom. We
are not making any holographic correspondence. We simply note that
the degrees of freedom reflect a theory of lower dimension, but we
do not explicitly postulate such a theory. In the case of a black
hole, our answer  agrees with that suggested by a stretched
horizon, although it would have some energy-dependent structure.
 Our procedure only works however in a curved background,
since it does not incorporate any back-reaction, although the
curved background does reflect strong gravitational effects.
Clearly, this is not sufficient to derive area-law scaling for all
systems. For example, we would have nothing to say with this
procedure about flat space. That is because what we are trying to
do is count only those states that can be properly treated with
low-energy field theory. States for which the back-reaction would
alter the gravitational background considerably should be
excluded. We do not exclude all such states so we will not always
see the Bekenstein bound. For example, although we eliminate all
cutoff sized black-holes, we clearly do not eliminate all possible
configurations that can form a black hole or have some other
strong gravitational back-reaction. Therefore, even for the curved
spaces we consider, we sometimes find certain parameter regimes
that reflect the full dimensionality of the space. Although it is
not the whole story,
 this simple procedure should
provide some guidance in seeking a more comprehensive
understanding of holography. It also suggests that even with the
full bulk description of the space, there is not necessarily
redundancy in the low-energy description, when the metric is
properly accounted for. It should also be kept in mind that
although we refer to a cutoff, we have in mind an ultimate
non-field theoretic description of the theory that kicks in at
that scale. The fact that our answers are cutoff dependent tells
us how the entropy associated with the cutoff should scale with
size of the system. Furthermore, because the local cutoff changes
with the curvature of the space, it suggests that degrees of
freedom that are best thought of as free fields in some contexts
are strongly bound states in others. Our work tells only about the
field theoretical contribution, which is in general less than the
full counting.

One advantage of our method of counting, even when there exists a
precise holographic description as in the AdS example, is that it
indicates the wave function for states in the bulk theory
associated with the holographic dual on the boundary. We will see
this explicitly for global AdS where we can see why the
Bekenstein bound applies to the bulk theory. Another advantage of
our method is that  regulates a theory to give finite entropy,
even in the presence of a horizon in a way well motivated by the
physics. 't Hooft previously had introduced a brick wall cutoff
to deal with this situation.

The organization of this paper is as follows. We will start by
reviewing the regularization we applied to RS1. In Section
\ref{secOTH}, we  present the techniques we use for counting
degrees of freedom or calculating entropy. We show
that the answer will be the answer one should expect,
the integrated local energy over the proper volume,
where only states consistent with the local cutoff are permitted.
We also demonstrate the
relation between our cutoff procedure and Pauli-Villars, which it
closely resembles.
Section \ref{secGADS} applies our methodology to
global anti-de Sitter space, where we again see the reduced dimensionality
from our simple procedure.
Sections \ref{secBH} and
\ref{secGDS} explore de Sitter space and black hole space-times.
We will see that the local UV cutoff obviates the need for
't Hooft's brick wall cutoff.
    In the
following section, we speculate about the non-covariant nature of
our result; in particular, why it might not entirely
account for all states in time-dependent
space-times.
Finally,
in section \ref{secCONC}, we summarize our results and their implications.

\section{Poincare Patch AdS \label{secPPADS}}

Before looking at the spatially-varying cutoff in general space-times, we
explain how it applies in the two brane RS1 model
\cite{rs1}. This model is fairly well understood,
phenomenologically viable, and believed to be holographic
\cite{phenomandhol,rz}. The background geometry is the Poincare
patch of 5D anti-de Sitter space with metric:
\begin{equation}
ds^2 = \frac{1}{(kz)^2}(-dt^2 + d{\mathbf x}^2 + dz^2) \label{rsmetric}
\end{equation}
The bulk cosmological constant is -$3k^2$ which we assume to have a
large magnitude, of order the Planck scale.  In RS, the AdS
horizon is cutoff by a Planck brane at $z_0 = \frac{1}{k} $ and a
TeV brane at $z_1 = \frac{1}{T} $, where $T$ is a mass scale
of order TeV. It is clear from
\eqref{rsmetric} that the induced geometry at any fixed $z$ is
flat. It is also not hard to show that at low energies the bounded
fifth dimension can be integrated out to get an effective theory
which is flat as well.

Let's suppose we
have a scalar field which is free to propagate in the bulk. A
simple hypothesis is that the total number of degrees of freedom
of such a field is given roughly by the volume of the space
normalized with some UV cutoff $\Lambda $ \cite{sussholo}:
\begin{equation}
g ( \Lambda )
\approx \Lambda^4 \int \sqrt{- G} d^3 x d z
\approx \Lambda^4 L^3 \int^{1/T}_{1/k} \frac{1}{( k z )^4} d z
\approx \Lambda^3 L^3 \Lambda ( \frac{k^4 - T^4}{k^5} )
\approx \Lambda^3 L^3 \frac{\Lambda}{k}
\label{volcalc}
\end{equation}
For large $k \approx \Lambda $, this looks like a 4D system.

Now, suppose we tried to count the degrees of freedom by adding up the
Kaluza-Klein (KK) modes of the bulk field. The massless field in 5D can be
decomposed into a set of 4D modes with masses given by roughly $m_n = j T$
for
integer $j$. Then each mode satisfies:
\begin{equation}
E^2 = p^2 + m_j^2
\end{equation}
where $p$ is the 3-momentum. The number of states with energy less than
$E$,
$g( E )$, is given by:
\begin{equation}
g ( E ) = \int_{ 0}^{E / T} d j L^3 \int_0^{\sqrt{E^2 - ( j T )^2}}
p^2 d p \sim E^3 L^3 \frac{ E}{T} \label{npp5D}
\end{equation}
So $g ( \Lambda ) =\Lambda^3 L^3 \frac{\Lambda}{T} $. This superficially
has the same form as \eqref{volcalc}, but we expect $\Lambda \approx k \gg
T$. Since
$\frac{1}{T}$ is the size of the fifth dimension, this wrongly
implemented Kaluza Klein picture
makes it seem like there are 5D degrees of freedom.

These two results are different because the volume calculation is
scaling the cutoff with position, while the KK calculation, as
presented above, is not. Indeed, the contribution to the volume at
a position $z$ is not $\Lambda^4 L^3 d z$ but $\frac{\Lambda^4
L^3}{k^5 z^5} d z$.

In \cite{us1,us2}, a position-dependent regulator was introduced. By
calculating Feynman diagrams in 5D, we showed that gauge
couplings in the 5D bulk run logarithmically, that is as in
4D, and that perturbative unification is feasible at
a high scale in $\mathrm{AdS}_5\;$. What we want to emphasize here
is not the details of the calculation, but why the result is to be
expected with a generally covariant cutoff in place.

When we did a field theory calculation in AdS$_5$, we did
not use explicit KK modes, but did our calculation in the five-dimensional
space with a mixed position space/momentum space formulation,
integrating over
$p_{\mu}$ and $z$. We applied a cutoff on momentum $p$ that varied
with position in accordance with the warp factor, that is
$p ( z ) < \Lambda $ or equivalently,
$p < \Lambda ( z )$ where
\begin{equation}
\Lambda ( z ) = \frac{1}{k z} \Lambda
\end{equation}
This is the spatially-varying cutoff on energy; the UV cutoff on $p$ depends
on position (see Figure 1). The highest cutoff is on the Planck brane, where
$\Lambda ( \frac{1}{k} ) = \Lambda$.
On the TeV brane, $p$ can only go up to $\Lambda \frac{T}{k} \ll \Lambda $. 
Because $p < $ $\frac{1}{k z} \Lambda $ is the same constraint as 
$z < \frac{\Lambda}{k p}$ and $p < \Lambda $, the position-dependent
cutoff is equivalent to a usual cutoff on  4D momenta and an
energy-dependent limit on $z .$ By taking this constraint on $z$
as the new boundary of the system, one can readily see that the
number of states is drastically reduced and agrees with the
holographic expectation.

\begin{figure}
\begin{center}
\psfig{file=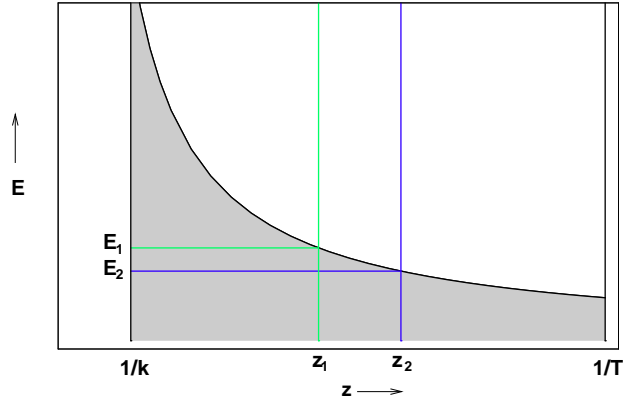,width=3.2in}
\end{center}
\caption{
In the RS1 model, an IR boundary at $z=1/T$ is effectively replaced for
energies $E>\frac{\Lambda}{k}T$. The new boundary, shown for
two modes with energies $E_1$ and $E_2$, is at
$z_{1(2)} = \frac{\Lambda}{k}\frac{1}{ E_{1(2)} }$. 
The area of the ($E,z$) plane 
satisfying $E<\Lambda(z)$ is shaded.}

\end{figure}

We emphasize that there are two aspects to this procedure. First,
there is a  spatial-cutoff for a given energy. Second there is a
re-quantization reflecting this spatial cutoff for each energy.
For this RS1 Poincare patch example, we get the right
counting even without moving the boundary.
However, in general, we have to
move the boundary explicitly for each energy to reproduce what we
expect from the more detailed mode analysis. This was the actual
procedure used in Ref. \cite{us1,us2}.

Not only does the above calculation demonstrate 4D behavior,
it shows that there is an intermediate regime that appears to be
5D \cite{phenomandhol}. The cutoff is position dependent only for 
$p > \frac{\Lambda}{k} T$. 
At lower energies, the cutoff is fixed at 
$z = \frac{1}{T}$. 
In this regime, the constraint on $z$ would have implied a brane deeper in
the IR than the TeV brane, so it is irrelevant. 
At very low energies $p < T$, no KK modes are excited and the theory is
4-dimensional: $g ( E ) = ( E L )^3$. 
If $T < p < \frac{\Lambda}{k} T$ then $E^2 =\mathbf{p}^2 + j^2 T^2$. 
Therefore in this
regime, one obtains the usual 5D dispersion relation,  and  
$g(E) =\frac{E}{T} ( E L )^3$. When $E/T > \Lambda / k$ we are 
in the holographic regime. In summary, the theory appears to be 4D 
for $E<T$, 5D for $\frac{\Lambda}{k} T>E>T$ and 4D for
$E>\frac{\Lambda}{k} T$.

\section{Other Geometries \label{secOTH}}

Now that we have seen how the spatially-varying cutoff leads to holographic
thermodynamics and $\beta$-functions in the RS1 model, we
generalize  to other geometries.
For time-independent metrics, the killing
vector $\partial_t$ allows us to assign an energy $E$ to any state. As we
justified above, this energy should never be greater than the locally
measured cutoff $\Lambda ( r ) = \sqrt{g_{t t} ( r )} \Lambda$.
The state with energy $E$ should
only probe the region where $\Lambda ( r ) \geqslant E$. So the boundary
$r_E$ at an energy $E$ is determined by:
\begin{equation}
E = \sqrt{g_{t t}( r_E )} \Lambda \label{elrel}
\end{equation}
The $E$-dependent cutoff on $r$ will determine the quantization of
momentum in the $r$ direction, and hence the quantization of $E$ itself.
In some cases, one can derive details about the spectrum; however, even
without the precise spectrum, we can evaluate approximately the density of states.

In the examples below, we focus on metrics which take the form
\begin{equation}
\mathrm{ds}^2\; = - V ( r ) \mathrm{d}t^2\; + \frac{1}{V ( r )}
\mathrm{d}r^2\; + r^2 d \Omega^2 \label{vmetric}
\end{equation}
Some examples are global anti-de Sitter space $V ( r ) = 1 + k^2 r^2$,
static de Sitter  $V ( r ) = 1 - k^2 r^2$, the exterior of
Reissner-Nordstrom black holes
$V ( r ) = 1 - \frac{2 m}{r} + \frac{q^2}{r^2}$.
This is an interesting class of metrics in that they are examples
in which area laws are known to apply. Furthermore, as we will
discuss, they all manifest UV/IR correspondence.

Consider the covariant Klein-Gordon equation:
\begin{equation}
\frac{1}{\sqrt{g}} \partial_{\mu} ( \sqrt{g} g^{\mu \nu} \partial_{\nu}
\phi ) = M^2 \phi  \label{ckg}
\end{equation}
We first consider states in the 4D system.
In the background \eqref{vmetric}, and with the ansatz
$\phi = \varphi ( r ) Y_{\mathrm{lm}\;} ( \theta , \phi ) e^{i E t}$,
\eqref{ckg} becomes:
\begin{equation}
\frac{E^2}{V ( r )} \varphi + \frac{1}{r^2} \partial_r ( r^2 V ( r )
\partial
_r \varphi ) - \frac{l ( l + 1 )}{r^2} \varphi - M^2 \varphi = 0
\label{qmV}
\end{equation}
This is just a 1D quantum mechanics problem. The spatially-varying cutoff
would enter
through
the boundary conditions which will depend on $E$.

With the exact spectrum, one can evaluate the density of states.
Even without the exact spectrum, one can often use the semi-classical (WKB)
approximation \cite{brick}. Alternatively, one can evaluate the density
of states based on the local energy and proper distance in accordance
with the metric. We will see that both these last two methods give yield the same
formula for calculating the density of states.

WKB applies when the phase of the wavefunction changes
much faster than the amplitude. We can then write $\varphi = \sqrt{\rho}
e^{i S ( r )}$ and assume that $k(r) = S' ( r )$ is large. This allows us to
solve \eqref{qmV} implicitly (for $M = 0$):
\begin{equation}
k(r,E,l) = \frac{1}{V ( r )} \sqrt{E^2 - \frac{V ( r )}{r^2} l ( l + 1 )}
\end{equation}
The number of oscillations of the phase of $\varphi$ over the whole space
is
the number of nodes of the approximate wavefunction. This is roughly the
number of modes with energy less than $E$. Indeed, as $E$ is lowered to
zero,
each of these nodes should disappear; whenever a node hits the endpoint,
there is another state. Thus the number of states with energy less than
$E$
is given by
\begin{equation}
g ( E ) = \int \mathrm{dr}\; \int k(r,E,l) ( 2 l + 1 ) d l = \int
\frac{\mathrm{dr}\;}{V ( r )} \int ( 2 l + 1 ) d l \sqrt{E^2 - \frac{V ( r
)}{r^2} l ( l + 1 )}
\end{equation}
The integral over $l$ is for values of $l$ which keep $k$ positive, that
is
$l( l + 1 ) < \frac{r^2 E^2}{V ( r )} $. This gives:
\begin{equation}
g ( E ) = \frac{2}{3} E^3 \int \frac{r^2}{V ( r )^2} d r \label{wkbint}
\end{equation}
Note that $g ( E )$ already seems to have 4D $E$ dependence.
So, unless the $r$ integral depends on $E$, we will have 4D thermodynamics.
In our regularization, we limit $r$ by $E < \sqrt{V ( r_E )} \Lambda$.
This will add $E$-dependence to the $r$ integral. For example,
if $V(r_0)=0$ at a horizon, then the integrand will have a pole. The only
modes that can probe up to the horizon have $E=0$, so the pole is an $E=0$
pole and must change
the $E$ dependence of the density of states for low energy. We will show
this explicitly in later sections.\\

In fact, \eqref{wkbint} is precisely the answer we expect
in four dimensions once the metric is properly accounted for. That is,
for an $n$-dimensional space, we expect the number
of degrees of freedom to be
\begin{eqnarray}
g(E) &\approx& \int^{r_E} \left(\sqrt{g^{tt}}E \right)^{n-1} r^{n-2}
\sqrt{g_{rr}} dr \\ \nonumber
& \approx& \int {E^{n-1} \over V(r)^{\frac{n}{2}} }r^2 dr
\end{eqnarray}

%
One can partially understand the  relation between the position-dependent cutoff and a
Pauli-Villars regulator
by examining the form of the equation of motion.
Since we are concerned with time-independent metrics, we write $\phi =
\varphi ( x ) e^{i E t}$ and the Klein-Gordon equation \eqref{ckg} becomes
\begin{equation}
g^{t t} ( x ) E^2 \varphi + \frac{1}{\sqrt{g}} \partial_i (
\sqrt{g} g^{i j} \partial_j \varphi ) = M^2 \varphi \label{kgt}
\end{equation}
The Green's function for a quantum field and for its PV
regulator at an energy $E$ will satisfy \eqref{kgt}. The PV field has
negative
norm so its propagator will cancel the regulated field wherever the PV
mass
has a negligible effect on the propagator. In particular, there will be a
thorough cancellation for all values of $x$ which satisfy
$g^{t t} (x) E^2 \gg M_{\mathrm{PV}}^2$.
This is equivalent to equation \eqref{elrel} for
$\Lambda = M_{\mathrm{PV}\;}$.

As an example of the equivalence between our regularization
procedure and Pauli-Villars, we look again at
the RS1 model. PV was used in this scenario in \cite{pom}.
The propagators for massless and massive fields
were derived in \cite{us1} and involve Bessel
functions
such as $\mathcal{J}_{\nu} ( p z )$ where $\nu = \sqrt{1 + M^2 / k^2} $
and $p$ is the momentum. These functions have the nice property that
$\mathcal{J}_{\nu} ( x )$ is independent of $\nu$ (up to a phase) for $x
\gg \nu$. Thus, the propagator for a field and its PV regulator will cancel
if
$z > \frac{\nu}{p} \approx \frac{M}{k p}$. This is exactly the condition
we use when applying the position-dependent cutoff.

Once we have $g ( E )$, we can generate all the important thermodynamic
quantities. The free energy is:
\begin{equation}
F = - \frac{1}{\pi} \int \frac{g ( E )}{e^{\beta E} - 1} d E \label{thermo}
\end{equation}
Which gives $U = \frac{\partial}{\partial \beta} ( \beta F )$ and 
$S = \beta (U - F )$.
For our examples, the scaling of $g(E)$ with $E$ will be the same
as the scaling of $S(T)$ with $T$.

\section{Global AdS \label{secGADS}}

In \cite{witten,hp}, the thermodynamics of global AdS was considered.
Witten found that at a temperature of order $k$ (the AdS
curvature scale), there is a transition from pure AdS to
AdS-Schwarzschild, above which both the bulk and boundary theory
reflect the number of degrees of freedom of a four-dimensional
theory in that the entropy scales as $T^4$. 
We now show that our counting
of states agrees with the above transition between
that of a low-energy
theory and one for which it reflects the full dimensionality (our
estimate of states will not reflect the gap).
More precise agreement would require choosing 
$\Lambda$  and the number of fields
in accordance with the holographic correspondence
(see below).

The AdS potential is $V(r)=1+k^2r^2$.
To apply our procedure, we introduce a boundary regulator
brane at a position $r=R$.
Also, for simplicity, we consider $AdS_5$ rather than $AdS_5\times S^5$.
As explained in the previous section, we expect the number of states
is approximately (note that now we are in 5D):
\begin{equation}
g(E) \sim E^4 \int_{r_E}^R {r^3 dr \over V(r)^{5/2}}
\sim E^4 \int_{r_E}^R {r^3 dr \over k^5 r^5} \label{ads5d}
\end{equation}
where the last approximation is valid for $r \gg 1/k$. In this
equation, $r_E$ is chosen in accordance with a spatially varying
cutoff as described in Section \ref{secOTH},
$\Lambda ( r ) = \sqrt{1 + k^2 r^2} \Lambda$. Setting $\Lambda ( r_E ) = E$ leads
to
\begin{equation}
r_E = \frac{1}{k} \sqrt{\frac{E^2}{\Lambda^2} - 1}\sim \frac{E}{k\Lambda}
\end{equation}
The number of states  is now readily evaluated to find
\begin{equation}
g(E) \sim \left( {E \over k} \right)^4 \left( {\Lambda \over E}
\right) \sim \left({E \over k} \right)^3 {\Lambda
\over k}
\end{equation}
This shows the dependence on energy of a four-dimensional
theory when $E$ is sufficiently large, as anticipated. Notice the
confinement of the high energy modes to the region near the
boundary yields the factor of $1/E$ that converts the behavior
from five to four-dimensional.

For low energies, $E<\Lambda$, the answer will not be that of a
lower-dimensional theory. That is to be expected, since for these
energies, one is probing distance scales less than $1/k$, for
which the theory should resemble flat space. This follows when
applying our regulator, since for energies less than $\Lambda$,
the entire space, down to arbitrarily small $r$, can support the
state. Actually, a more careful analysis with the precise modes
would reflect the band gap that would limit this 5D scaling
behavior to energies between $k$ and $\Lambda$.  In summary,
the theory appears to have   5-dimensional counting of states for
$k<E<\Lambda$ and appears 4-dimensional for $E>\Lambda$. Notice
that this behavior closely resembles what we found for the
Poincare patch calculation, which had the fewest modes at low
energies, then had a five-dimensional regime, then evolved to the
holographic four-dimensional regime.

We can ask what we learn from this method that we  did not already
know by using the precise holographic dual \cite{sw}. The answer
is that we have a guide to the precise nature of the
correspondence, since we can see what states exist in the bulk
dual theory. For example, in the state counting done by Susskind
and Witten, the Bekenstein bound was assumed for the bulk.
But if we use the additional information of precisely what the
cutoff should be as determined by the duality, we can see this
counting of bulk states explicitly.  To see this, we use the
coordinate parameterization of AdS space assumed in that paper,
where $V(r)=1/(1-(kr)^2)^2$ and $k$ sets the AdS curvature
scale. Then with a regulator brane  a distance $\delta$ from the
boundary, the 4D theory would have a cutoff on distance $\delta$
so that  the cutoff on proper distance is $1/k$. Then the volume of
the AdS$_5$ is 
$(1/k) \times {\mathcal A} \times k^4$, where ${\mathcal A}$ is the 3D
area and we have used the cutoff $k$ provided by the dual holographic
theory, which we know for this example. 
From this, we almost have a result well within the Bekenstein bound. 
But this was for a
single field in the bulk. We expect
that the description of the holographic dual would require $N^2$
such fields so that the total maximum entropy would be $N^2 {\mathcal A}k^3$,
where $N$ is a parameter from the dual theory.
\footnote{We thank Massimo Porrati for discussions of this calculation.}

We can also obtain the answer for the density of states
through studying directly the
eigenvalue spectrum. For simplicity, we show a 4D AdS example,
rather than the 5D example we just looked at.
When $E > \Lambda $ we restrict $r$ by
$r_E < r < R$. Let us take the limit $R k \gg 1$ and $E\gg\Lambda$.
If we change variables to $\rho =
\frac{r_E}{r} $ and define $\varphi = \rho \chi ( \rho )$ then in
this limit, \eqref{qmV} simplifies to an analog quantum mechanics
problem:
\begin{equation}
- \partial^2_{\rho} \chi + \frac{2}{\rho^2} \chi -
\frac{\Lambda^2}{k^2} ( 1 - \frac{l ( l + 1 ) k^2}{E^2} )
\chi = 0
\end{equation}
Note that we have absorbed the spatially-varying cutoff into the
normalization of $\rho $ so this equation has boundary conditions
at fixed $\rho $. The solutions are Bessel functions (in fact,
they are the same Bessel functions as in the 4D Poincare patch
case). The eigenvalues for each $l$ are integers $j_l$ related to
the energy and other parameters as:
\begin{equation}
j_l^2 = \frac{\Lambda^2}{k^2} ( 1 - \frac{l ( l + 1 )
k^2}{E^2} )
\end{equation}
In other words:
\begin{equation}
E^2 = \frac{l ( l + 1 ) k^2}{1 - ( \frac{n k}{\Lambda} )^2}
\end{equation}
Therefore the number of modes at each $l$ is bounded by $n <
\frac{\Lambda}{k}$.
The total number of modes less than $E$ is $g ( E )\sim
\frac{\Lambda}{k} ( \frac{E}{k} )^2$. This density of states
is 3D which is holographic to the 4D background.

Notice that the energy scale here is different
from the Poincare Patch example.
With the regulator brane there is the potential for a phenomenologically viable
4D theory of gravity if we choose the energy cutoff to be of order $M_{Pl}$,
which is related to the 5D Planck scale $M$ by $M_{Pl}=(\frac{M}{k})^{3/2}k^2 R$.
For large $R$ this is a much higher cutoff than
the natural expectation $\Lambda\approx M$ which we have used.

\section{Black Hole Thermodynamics \label{secBH}}

In this section, we  consider the contribution to the entropy
from a scalar field in the exterior of a black hole
\cite{brick,myers}. Although this is  not the fundamental
contribution to a black hole's entropy, it can give us information
about the scaling with cutoff and dimension of this fundamental
contribution which
should have the same dimension-dependence.

Black holes and de Sitter space (see next section) differ from the
AdS examples we just studied in that states are concentrated not
on a boundary where $g_{tt}$ is largest but on a horizon, where it
is smallest. It might seem surprising that our method of counting
states would yield a concentration of states on the horizon,
since, as we have emphasized, the number of states is reduced by
restricting states to a region where they are consistent with the
local cutoff. By this reasoning, we would expect high energy
states to be concentrated away from the horizon so that the
entropy would also be concentrated there.

However, state counting proceeds differently in the black hole and
de Sitter space examples, because we assume that there is a fixed
temperature, $T$, as measured at infinity (or the origin for de Sitter space).
Therefore, in general, the energy stays well
below the local cutoff. In fact, energy at a given position $r$
will be chiefly of order $T/\sqrt{V(r)}$, where $V(r)$, given in
Section \ref{secOTH}, goes to zero near the horizon. This means that for a
fixed $T$, the highest energy region is where $V(r)$ is smallest,
the opposite of what happens if we allow all energies up to the
cutoff at all $r$. In fact, energies only achieve the local
cutoff near the horizon. With no cutoff on energy, they would
diverge as one approaches the horizon. We assume a cutoff on
energy $\Lambda$ which we expect to be of order $M_{Pl}$. This in
turn implies a cutoff on position; states cannot get too close to
the horizon unless they have arbitrarily low energy (as measured
at infinity).

For a black hole,
$V(r)=1 - \frac{r_S}{r}$, where $r_S = \frac{2
m}{M^2_{Pl}}$ is the location of the horizon and $m$ is the mass
of the hole. Generalizations to rotating or charged black holes
are straightforward. The unregulated density of states diverges at
$r = r_S$ because $V ( r_S ) = 0$. Following the prescription of
Section \ref{secOTH}, we define the cutoff to be $\Lambda $ at $r
= \infty$. Then the closest a state of energy $E$ can get to the
horizon is determined by $\sqrt{V ( r_E )} \Lambda = E$. So,
\begin{equation}
r_E = r_S \frac{\Lambda^2}{\Lambda^2 - E^2} \label{rebh}
\end{equation}
The solution to the analog quantum mechanics problem has been
extensively studied without energy-dependent boundary conditions
(see, for example, \cite{sskk}).  For our
purposes, it will be sufficient to estimate the density of states
using Section \ref{secOTH}.

In order to regulate the IR divergence from the asymptotic
Minkowski space, we restrict space to  a box of size $L$.  Then
the number of states calculated with  \eqref{rebh} and \eqref{wkbint} is
\begin{equation}
g ( E ) \approx \frac{2}{3} r_S^3 \Lambda^2 E + \frac{2}{9} L^3 E^3
+\mathcal{O} ( \frac{r_S}{L} ) + \cdots \label{geus}
\end{equation}
The $L^3 E^3$ term is just what we expect from flat space. The
other term is the leading contribution associated with  the black hole. The
cutoff dependence comes from a factor of $r_E-r_S$, which scales as
$(E/\Lambda)^2$. The entropy which follows from the
$L$-independent part of this $g ( E )$ scales with temperature as:
\begin{equation}
S ( T ) \propto r^3_S T \Lambda^2 +\cdots \label{stus}
\end{equation}
Substituting in the Hawking temperature, we see the black
hole contribution scales with the horizon area.

If we used the Reissner Nordstrom black hole potential,
$V(r) = (1-\frac{r_+}{r})(1-\frac{r_-}{r})$ \eqref{stus} would
have been replaced by
\begin{equation}
S ( T ) \propto \frac{r^4_+}{r_+-r_-} T \Lambda^2 + \cdots  \label{stusrn}
\end{equation}
Note that the RN temperature is $T_{RN} = \frac{r_+-r_-}{4\pi r_+^2}$ which makes
$S(T_{RN})\propto {\mathcal{A}}\Lambda^2$ where ${\mathcal{A}}=4\pi r_+^2$ is the
area of the horizon.

This computation follows closely that performed by 't
Hooft in Ref. \cite{brick}, in which he assumed a sharp cutoff at
a coordinate $h$ corresponding to a proper distance of order
$1/M_{Pl}$.  In our approach,
states are kept away from the horizon due to a cutoff on energy,
so that a state with finite energy cannot reach the horizon. So
the minimum distance of a state from the horizon depends on
energy. Because the energy at a position $r$ is of order
$T/\sqrt{V(r)}$, the contribution is heavily concentrated in the
high proper energy states that are closest to the horizon. In fact, it is
easy to see that the proper distance of these states from the
horizon is of order $1/M_{Pl}$, as in 't Hooft's calculation.

Our calculation is even closer in spirit to that of Demers, Lafrance, and Myers
\cite{myers}.
They also calculated the entropy contributed
from a scalar field external to a black hole. Their calculation
employed a Pauli-Villars regulator, which we have already shown
closely matches our regulator. They
furthermore verified the suggestion of Susskind and Uglum Ref. \cite{susGN} by
demonstrating the renormalization of the entropy was consistent.
In flat space, Newton's constant gets renormalized as:
\begin{equation}
\frac{1}{G_N} \rightarrow \frac{1}{G_N} + \frac{B}{12 \pi} \label{rengn}
\end{equation}
where $B$ is some quadratically divergent function of the five PV masses.
Then they use the same PV fields to regulate the divergence of the entropy outside a
black
hole. The entropy they get, using the same WKB technique, is
$S ( T ) \propto \frac{r_+^4}{r_+-r_-} T B$.
Note the similarity to \eqref{stusrn}. They then interpret this as a
renormalization of the black hole entropy:
\begin{equation}
S_{\mathrm{BH}\;} = \frac{\mathcal{A}}{4 G_N} \rightarrow
\frac{\mathcal{A}}{4
} ( \frac{1}{G_N} + \frac{B}{12 \pi} )
\end{equation}
The important point is that this is the same
quadratically divergent function $B$ as in \eqref{rengn}.

It is also straightforward to work out the logarithmic corrections.
We obtain a term in \eqref{stusrn} proportional to
$\frac{r_+^4}{r_+-r_-}T^3 \log(\Lambda)$. 
This piece corresponds to the divergent contribution
$T^3\log(M)$ in \cite{myers}. The scale for the logarithm
is naturally set by the black hole temperature $T_{RN}$.
We can interpret this logarithm as a
constant contribution related to higher curvature terms on the
gravitational action \cite{myers,higher}, being absorbed in the
renormalization of higher dimension operators in the gravity action.

This result would be especially appealing in a theory of induced
gravity.
There the fact that both $G_N$ and the entropy receive the same
contribution which grows with the number of degrees of freedom
would guarantee there is no problem with a large number of
species. We also observe that in this case, one might be able to
understand the small value of $G_N$ as a result of a large number
of species.

Clearly, there is a contribution to black hole entropy not
associated with an external field. Very likely, this contribution
is associated with whatever regulates the quantum field theory at
energies of order the cutoff $\Lambda$. From this perspective, it
is not surprising that there exists a string theory example
where the full entropy is reproduced \cite{stromvafa}. Furthermore, this shows
on quite general grounds that whatever the cutoff physics
contribution is, it
should scale quadratically with associated mass scale and be
proportional to area, in order to absorb the cutoff dependence of
our field theory result, rendering it scheme-independent.

Finally, it is of interest to reflect on the form of the result.
 We found that for a black hole at the Hawking temperature, the
entropy had a term that scaled with the volume of the space and an
additional term that is attributed to the black hole that scales
as $S \propto \Lambda^2 R^2$, where $\Lambda$
is the cutoff and $R$ is the Schwarzschild radius. In fact, the
answer really scales as $E^3 R^3 (\Lambda/E)^2$, and when
averaging over the thermal ensemble, $E$ gets replaced by $T$,
which is taken to be the Hawking temperature. Clearly, the first
factor is what one would expect for the volume inside the black
hole horizon. From this perspective, we see that the number of
degrees of freedom is in fact far {\it greater} than the expected
number for a system at such low temperature. How are we to
understand this result?

Let us consider the value of $r_H-r_S$, where $r_H$ is the minimum
$r$ permitted by our regularization procedure for a mode of energy
$T_H=1/R$. One finds $(r_H/r_S-1)=(T_H/\Lambda)^2$.  This
corresponds to the {\it proper} distance from the horizon equal to
$1/\Lambda$, what one would expect to be the minimum distance
permitted in a system with cutoff $\Lambda$.  Notice that in this
sense, our regularization is naturally leading to the notion of a
stretched horizon \cite{sussthor}. The degrees of freedom are concentrated
at and around the coordinate $r_H$. Of course, we are not making
any further claims; we do not consider a membrane as a physical
object. We are simply making the observation that with our cutoff
procedure, the degrees of freedom naturally tend to congregate at
this position.

Furthermore, we understand the enhancement of the number of
degrees of freedom over that which is expected. We see that the
proper energy at $r_H$ is actually $\Lambda$, not the much smaller
$T_H$.  This is not surprising as this is how $r_H$ was chosen.
But we see that the large number of degrees of freedom is due to
the much higher proper energy of the modes at $r_H$.  In fact, we
can eyeball the result by taking the quantity $E_p^3 \delta
r_p^3$, where $E_p=E/(\sqrt{r_h/r_S-1})$ is the proper energy and
$\delta r_p=(r_h-r_S)/(\sqrt{r_h/r_S-1})$. Of course, this is
restating the interpretation given in general in Section \ref{secOTH}.

\section{Static Patch of de Sitter Space \label{secGDS}}

The final example we consider is de Sitter space, for which the
entropy is bounded and is expected to take the form
$S_{\mathrm{BH}\;} \propto M^2_{\mathrm{Pl}\;} R^2 .$ We can
compute the quantum corrections to this formula, as above for the
exterior of the black hole.

The de Sitter function is $V ( r ) = 1 - \frac{r^2}{R^2} $ with $0
< r < R$. Since $V ( r )$ decreases as we approach the boundary,
the cutoff $\Lambda ( r ) = \sqrt{V ( r )} \Lambda $ has its
maximum at $r = 0$. Setting $\Lambda ( r_E ) = E$ gives $r_E = R
\sqrt{1 - \frac{E^2}{\Lambda^2}} $. The state counting of Section
\ref{secOTH} gives:
\begin{equation}
g ( E ) = \frac{R^3}{3} \Lambda^2 E \left( \sqrt{1 - \frac{E^2}{\Lambda^2}}
-
\frac{E^2}{\Lambda^2} \mathrm{Tanh}^{- 1}\; ( \sqrt{1 - \frac{E^2}{\Lambda^2}} ) \right)
\end{equation}
This function looks like a hump peaked around $E \approx 0.46
\Lambda $.

However, we are only interested in counting the states at low
energy, since we  assume the space is at the de Sitter
temperature. Then,
\begin{equation}
g ( E ) =  \frac{R^3}{3} \Lambda^2 E + \cdots \label{dglowe}
\end{equation}
The linear energy-dependence arose because the distance to the horizon of
a state of energy $E$ scales as $(E/\Lambda)^2$, as in the black
hole example, which has the same near horizon structure.

The entropy is
\begin{equation}
S \propto R^3 \Lambda^2 T
\end{equation}
Evaluated at the de Sitter temperature
$T = T_{\mathrm{dS}\;} \approx \frac{1}{R}$ this says that
$S_{\mathrm{dS}\;} = R^2 \Lambda^2$. We expect that as with the
black hole entropy, there is an associated renormalization of
Newton's constant. There are also logarithmically divergent
pieces which correspond to the renormalization of higher dimension
operators, as with black holes.

Notice that as with the black hole case, this is a much {\it
larger} number of states than would be expected in a box of radius
$R$ in flat space at temperature $T$, which scales like $R^3 T^3$.
This enhancement
arises in a similar manner to the enhancement of states for a
black hole at low temperature that we already discussed.
Notice also that the entropy we calculated is not obviously a restriction
on the number of states, but just corresponds to the number of states
at the de Sitter temperature. The fundamental description of the
system in principle involves many more states, most of which do not
participate.

\section{Coordinate Dependence}
In the previous sections, we illustrated the utility of the
spatially varying cutoff in reproducing qualitatively the known
results for the entropy in several different geometries.  The
AdS case is well understood, although there are still puzzling
features such as the explicit implementation
of the UV/IR correspondence \cite{polsussto}. The black hole case is
less well understood and exhibits some interesting features
that we discuss in more detail in this section.

Although we understood the form of our answer in terms of local
energy and proper volume,  there is an obvious issue that we have
so far skirted which is the coordinate dependence of our result.
Some of the coordinate dependence of course comes from the
non-covariant nature of the way we implemented our cutoff. A more
careful analysis would use Pauli-Villars, which we have shown our method
closely approximates. We use our procedure because the counting is
very simple.

However, more worrisome for the black hole example is that one
can choose coordinates that are completely smooth at the horizon.
Where would the cutoff dependence of black-hole entropy then
arise? We will necessarily be speculative in our considerations in
this section. However, we find it suggestive that our results
overlap with proposals that have already been made.

The key criterion that distinguishes the metrics in which we do
our calculations from those we have avoided is that the ones we
use are {\it time-independent}. For metrics that are strongly
time-dependent, our cutoff procedure simply does not apply and we
do not know how to do a calculation of the sort we have outlined
since we cannot count energy eigenstates as we have been doing. We
will next explain what goes wrong in several alternative
parameterizations of the black hole metric. We will also argue by
following the details of the coordinate transformation that  there
might exist long wavelength modes or nonlocal correlations that
are not accounted for in low-energy field theory and which are
sensitive to the cutoff physics. 
This would be a correction to  low-energy
field theory when counting degrees of freedom. However, these
modes should be very low energy and long wavelength, and therefore
irrelevant at a practical level, in which field theory should
still apply.

\subsection{Kruskal Coordinates}

To get coordinates which are smooth at the horizon, we first transform from Schwarzschild
($r$,$t$) to advanced and retarded Eddington-Finkelstein coordinates:
\be
U &=& t+r - r_s\log|r/r_s-1|\\
V &=& t-r - r_s\log|r/r_s-1|
\ee
The metric in these coordinates is still not well-behaved at the horizon, but it remains
Minkowski for large $r$. In Kruskal coordinates:
$x=e^U$, $y=e^V$, the metric is:
\be
ds^2 = {32 M^3 \over r} e^{-r/r_S} dx dy + r^2 d \Omega^2
\ee
There is no longer a singularity at the horizon. However for large $r$, the metric
looks like $ds^2 = \frac{dx dy}{xy}+r^2d\Omega^2$.
This is not Minkowski and one has to worry
about physics far away from the horizon.
As an example of how we must be careful using our physical intuition in these coordinates,
consider a mode near the horizon. It has very small energy measured at infinity,
and we define $\epsilon = \frac{E}{\Lambda}$. The distance to the horizon
this mode can probe is $r_{min} = r_S +\epsilon^2 r_S$.
To measure this mode, one would need an
amount of Schwarzschild time approximately equal to
$\Delta t = \frac{1}{E}=\frac{1}{\epsilon\Lambda}$. In Eddington-Finkelstein coordinates
this means that
$\Delta U \approx \Delta V \approx \frac{1}{\epsilon}$. While
this is large, the space-like combination $U-V$ is still small.
However, in Kruskal coordinates,
$\Delta x \approx \Delta y \approx \epsilon e^{\frac{1}{\epsilon \Lambda}}$.
Now the distance measured with respect to the space-like combination
$x-y$ is exponentially large. Since the space is flat near the horizon,
this corresponds to an exponentially large proper distance as well.

Tidal effects make any given region
grow with $t$ in terms of $x$ and $y$ coordinates.  We
see that taking the minimum time that would be necessary to probe low
energy states near the horizon, the near-horizon region is
transformed into a huge region. We would expect a large number of
low-energy states associated with this patch.
These are not anticipated in the field theory
associated with the patch that corresponded to the near-horizon
region.

It is of interest to consider the possibility of such low-energy
states.  Because they are very large and low-energy, they would
not affect any local physics so they are not precluded by the
success of low-energy field theory.  In fact, they are not
necessarily states but could be correlations in existing states
that store information. In this sense, they would be similar to
the ``precursors'' suggested in Ref.  \cite{polsussto,st}. Also, because these
states are sensitive to the cutoff of the theory, as has already
been emphasized, we see this would implement a UV/IR
connection \cite{sw,peet}. Finally, it is perhaps not unexpected that such
large nonlocal states should exist in this new coordinate system,
since the existence of an event horizon summarizes physics that is
extended in time. In the new coordinates, where time and space are
mixed, there might be new extended states.

Although we did the exercise for Kruskal coordinates, we expect
similar results for any time-dependent coordinate system. One
might expect things to simplify in coordinates which are well-behaved at the horizon
but smoothly approach Minkowski space far away. We
briefly consider one family of such coordinates, nice slice,
in the next section and demonstrate that the situation is just as
confusing.

\subsection{Nice Slice}
By Birkhoff's theorem, we know that the only set of coordinates in which
the black hole
metric is time-independent and asymptotically Minkowski is
Schwarzschild coordinates.
Nevertheless, we can go to a set of coordinates which are
nonsingular near the horizon and have a sort of minimal time dependence
\cite{polch}.
We define a one parameter ($R$) family of slicings implicitly by:
\begin{equation}
-e^{T/2r_S}y + e^{T/2r_S}x = 2R \label{ns}
\end{equation}
$x$ and $y$ are Kruskal coordinates defined above.
$R$ can take values from $0$ to $2 r_S$.
For $R=0$, $T$ is just Schwarzschild time $t$. So, by varying $R$ we can
introduce time-dependence in a controlled way.
To get the metric, we can solve \eqref{ns} for $T$
and then solve for a coordinate $Z$ orthogonal with respect to the black
hole metric. The solution is:
\begin{eqnarray}
T &=& -2r_S \log\left(\frac{R+\sqrt{R^2+4 r_S^2x y}}{2 r_S x}\right) \\
Z &=& -2\sqrt{R^2+4 r_S^2 x y}
      + 2 R \log\left(\frac{R+\sqrt{R^2+4 r_S^2 x y}}{2 r_S x}\right)
\end{eqnarray}
Note that $\sqrt{R^2+4 r_S^2 x y}=\frac{1}{2 r_S}(R T +r_S Z)$ which lets
us express $xy$ in terms of $RT+r_S Z$, both of which are independent of
Schwarzschild time $t$.
The metric in nice-slice coordinates is:
\begin{equation}
ds^2 = \frac{ r_S }{ 4 r }e^{-\frac{r}{r_S}}
\left(-\frac{(R T + r_S Z)^2}{4 r_S^4}dT^2 + dZ^2\right)
\end{equation}
Although $\partial_t$ is still a killing vector, this metric does
depend on the new time parameter $T$. For any $R$, $T \ra t$ as
$r \ra \infty$. For $R=0$, $T$ becomes $t$ everywhere, but the
metric becomes singular at the event horizon $r=r_S$
($RT + r_SZ=2r_S R$).
For nonzero
$R$ space is flat (by construction) at the event horizon, but the
coordinate singularity has moved to $RT+r_S Z=0$. This is a
surface of constant $r$ solving $\frac{R^2}{4r_S^2} =
e^{r/r_S}(1-\frac{r}{r_S})$. For $R=2r_S$, the horizon moves to
the essential singularity at $r=0$. In some sense, all we are
doing is moving the coordinate singularity between the event
horizon and the physical singularity by varying $R$.

Now suppose we tried to do quantum field theory in such a
background. Again, we would have to worry about states within
$\epsilon$ of the horizon. In nice slice coordinates, the position
of the horizon depends on time as depends on time as
$Z\approx\frac{2Rt}{r_S}$. Since these modes have time uncertainty
$\Delta t=\frac{1}{\epsilon \Lambda}$ we get that $\Delta Z\approx
\frac{R}{\epsilon \Lambda}$. So while in Schwarzschild
coordinates, the mode is localized within $\epsilon$ of the
horizon, in nice slice the mode is not localized at all. That is,
while the horizon is just a line in nice slice, the near horizon
region is very very large. If we tried to regulate the theory, we
would have to include finite time in our regulator to somehow take
this into account. Clearly this is a peculiar situation and not
something we really understand. It might also be connected with
the existence of new, low-energy states or nonlocal correlations.

\section{Conclusions \label{secCONC}}
Let us summarize our method and its justification.
We define a space-dependent condition on energy, or equivalently
a cutoff on local energy. By interpreting this as a cutoff on position,
we find an energy-dependent boundary for our space which is used to
quantize the system and evaluate the density of states.
One can also use the space-dependent cutoff to do field theory,
as in Ref. \cite{us1,us2}.  Another application is to testing
qualitative features of gravitational theories. For example,
it is readily seen by a field theory calculation
with our regulator that the proposal of Ref. \cite{thomas} to address
the cosmological constant solely through the holographic nature of the theory
does not work. For example, in the AdS case, all the degrees of freedom
are concentrated where the energy is largest, with no additional
red-shift factor.

Our counting
relies on the fact that any state above the local cutoff cannot
be treated simply in field theory. Above the cutoff, one expects
to find either black holes or states intrinsic to the fundamental
theory providing the cutoff, e.g. string theory. At short
distance, these are not weakly interacting field theory states.
For this reason, the standard argument, relying on a UV fixed
point,  that a boundary theory must reflect an asymptotically AdS
space does not apply.

In addition, the bulk theory description that we have given
suggests additional ``stored'' degrees of freedom that have become
strong bound states at the cutoff. For example, a rolling scalar
field in AdS space might change the vacuum energy so that many
more or fewer states appear in the region that appears
non-holographic from a field theory vantage point. These states
must be present once the curvature changes; do they emerge out of
nowhere or is it that they have dropped from above to below the
cutoff? So we interpret the entropy bounds as bounds on the degrees
of freedom that can be simultaneously excited in practice. 

We do not assume the existence of a boundary theory.
However, in all cases we have studied with a monotonic $g_{tt}$
that varies sufficiently strongly, we find the degrees of freedom
concentrated on the boundary of the space.
It is not clear that there exists a more useful boundary description in
general.

We also have seen that this approach is coordinate dependent.
However, for most of our examples, there is a unique choice of
metric for which there is not time-dependence. If there is a
gauge-invariant formulation, our approach should be the result of
a particular gauge choice.  It seems clear that time-dependent
metrics are more subtle to understand. Despite the existence of a
covariant formulation of the holographic principle \cite{bousso},
it is not clear how to apply any of the defining quantities we
have used, namely energy eigenstates, temperature, and entropy in
a strongly time-dependent vacuum.

That low-energy states or non-local correlations can exist and be
consistent with all known field theory successes is important.
Local measurements would not know about big low-energy states
since they carry low energy and their effects would be suppressed
by small wave function overlap, or equivalently, a large
normalization factor. However, if they are present, they can
provide correlations that can in principle be measured over large
distances. Such non-local effects should be relevant to the
information problem of black holes \cite{page}.

It is  interesting to contrast our approach with the ``gauge
theory" approach that has been suggested by which one would have
some principle through which one could eliminate redundant degrees
of freedom.  As we have said, our procedure
depends on a time-independent coordinate choice.
It does not however  preclude existence of a ``gauge" theory where
this is more readily embodied in a covariant way;
in such a formulation our answer should
correspond to a choice of gauge.

Of course, there is much more to fully understanding the
Bekenstein bound and holography than the simple procedure we have
outlined. However, it does well approximate the counting of states
in spaces that are highly curved even without exciting additional
fields. Also it might serve as a limit or special case of a more
general more covariant formulation.

The reason our procedure does not supply the final answer is that
we have not incorporated any back-reaction and we have not
supplied the physics of the cutoff in our counting. However,
it is a remarkably simple procedure to use in order to deduce
qualitative features of a gravitational system.  If we could
also exclude extended (large) black holes, we might
always find area-law behavior. The problem is that this involves a
constraint
on
both energy and size; there is not way to do this in a
conventional field theory. However, better understanding this
constraint might yield some insight into UV/IR.

\section{Acknowledgements}
We would like to thank Nima Arkani-Hamed, Lubos Motl, 
Joe Polchinski, Massimo Porrati, Steve Shenker, Andy Strominger, and Nick Toumbas.
V.S. would like to thank C. Nunez for enlightening
discussions and and J. Edelstein, N. Rius and M. Schvellinger for useful comments.
This work is partially supported by the Spanish MCyT grant PB98-0693.


\begin{thebibliography}{99}

\bibitem{thholo}
G.~'t Hooft,
arXiv:hep-th/0003004.

\bibitem{sussholo}
D.~Bigatti and L.~Susskind,
arXiv:hep-th/0002044.

\bibitem{maldacena}
J.~Maldacena,
Adv.\ Theor.\ Math.\ Phys.\  {\bf 2}, 231 (1998)
[Int.\ J.\ Theor.\ Phys.\  {\bf 38}, 1113 (1998)]
[arXiv:hep-th/9711200].


\bibitem{witten}
E.~Witten,
Adv.\ Theor.\ Math.\ Phys.\  {\bf 2}, 253 (1998)
[arXiv:hep-th/9802150].

\bibitem{gubser}
S.~S.~Gubser, I.~R.~Klebanov and A.~M.~Polyakov,
Phys.\ Lett.\ B {\bf 428}, 105 (1998)
[arXiv:hep-th/9802109].

\bibitem{polsussto}
J.~Polchinski, L.~Susskind and N.~Toumbas,
Phys.\ Rev.\ D {\bf 60}, 084006 (1999)
[arXiv:hep-th/9903228].

\bibitem{ldt}
O.~Aharony, M.~Berkooz, D.~Kutasov and N.~Seiberg,
JHEP {\bf 9810}, 004 (1998)
[arXiv:hep-th/9808149].

\bibitem{us1}
L.~Randall and M.~D.~Schwartz,
JHEP {\bf 0111}, 003 (2001)
[arXiv:hep-th/0108114].

\bibitem{us2}
L.~Randall and M.~D.~Schwartz,
Phys.\ Rev.\ Lett.\  {\bf 88}, 081801 (2002)
[arXiv:hep-th/0108115].

\bibitem{rs1}
L.~Randall and R.~Sundrum,
Phys.\ Rev.\ Lett.\  {\bf 83}, 3370 (1999)
[arXiv:hep-ph/9905221].


\bibitem{phenomandhol}
JHEP {\bf 0108}, 017 (2001)
[arXiv:hep-th/0012148].

\bibitem{rz}
R.~Rattazzi and A.~Zaffaroni,
JHEP {\bf 0104}, 021 (2001)
[arXiv:hep-th/0012248].

\bibitem{brick}
G.~'t Hooft,
Nucl.\ Phys.\ B {\bf 256}, 727 (1985).

\bibitem{pom}
A.~Pomarol,
Phys.\ Rev.\ Lett.\  {\bf 85}, 4004 (2000)
[arXiv:hep-ph/0005293].

\bibitem{sw}
L.~Susskind and E.~Witten,
arXiv:hep-th/9805114.





\bibitem{hp}
S.~W.~Hawking and D.~N.~Page,
Commun.\ Math.\ Phys.\  {\bf 87}, 577 (1983).

\bibitem{sskk}
D.~G.~Boulware,
Phys.\ Rev.\ D {\bf 11}, 1404 (1975).

\bibitem{myers}
J.~G.~Demers, R.~Lafrance and R.~C.~Myers,
Phys.\ Rev.\ D {\bf 52}, 2245 (1995)
[arXiv:gr-qc/9503003].

\bibitem{susGN}
L.~Susskind and J.~Uglum,
Phys.\ Rev.\ D {\bf 50}, 2700 (1994)
[arXiv:hep-th/9401070].

\bibitem{higher}
T.~Jacobson and R.~C.~Myers,
Phys.\ Rev.\ Lett.\  {\bf 70}, 3684 (1993)
[arXiv:hep-th/9305016].
R.~M.~Wald, Phys.\ Rev.\ D {\bf 48}, 3427 (1993)
[arXiv:gr-qc/9307038].
M.~Visser, Phys.\ Rev.\ D {\bf 48}, 5697 (1993)
[arXiv:hep-th/9307194].


\bibitem{stromvafa}
A.~Strominger and C.~Vafa,
Phys.\ Lett.\ B {\bf 379}, 99 (1996)
[arXiv:hep-th/9601029].

\bi{sussthor}
L.~Susskind, L.~Thorlacius and J.~Uglum,
Phys.\ Rev.\ D {\bf 48}, 3743 (1993)
[arXiv:hep-th/9306069].

\bibitem{st}
L.~Susskind and N.~Toumbas,
Phys.\ Rev.\ D {\bf 61}, 044001 (2000)
[arXiv:hep-th/9909013].

\bibitem{peet}
A.~W.~Peet and J.~Polchinski,
Phys.\ Rev.\ D {\bf 59}, 065011 (1999)
[arXiv:hep-th/9809022].


\bibitem{polch}
J.~Polchinski,
arXiv:hep-th/9507094.

\bibitem{bousso}
R.~Bousso,
JHEP {\bf 9907}, 004 (1999)
[arXiv:hep-th/9905177].


\bibitem{thomas}
S.~Thomas,
arXiv:hep-th/0010145.

\bibitem{page}
D.~N.~Page,
arXiv:hep-th/9305040.

\end{thebibliography}
\end{document}